\documentclass[12pt]{article}
\usepackage{pic02}
\usepackage{hyperref}
\usepackage{url}
\usepackage{graphicx}

\begin{document}

\title{\bf THE ANOMALOUS CHROMOMAGNETIC DIPOLE MOMENT OF THE TOP QUARK IN DIFFERENT FRAMEWORKS}
\author{R. Martinez and J-Alexis Rodriguez \\
{\em Departamento de Fisica, Universidad Nacional de Colombia
Bogota, Colombia} \\}
\maketitle

\baselineskip=14.5pt
\begin{abstract}
We give explicit
formulae for the anomalous chromomagnetic dipole moment of the top quark in the framework of the standard model,
the two higgs doublet model and the minimal supersymmetric standard model.
Finally, we compare the results for this coupling with the bound $-0.03 \leq \Delta
\kappa \leq 0.01$ coming from analysis of $b \to s \gamma$ process whith an on-shell gluon.
\end{abstract}

\baselineskip=14.5pt
The top quark is the heaviest fermion in the standard model (SM) with a mass of
$174 \pm 5.2 $ GeV.  Anomalous couplings  between
top quark and gauge bosons might affect top quark production and its decay at high energies. On the other hand, the 
$\bar{t} t g$ coupling correction to the
total top quark production cross section at Fermilab was calculated by
Stange et. al. \cite{willen}, in the framework of the SM. Since the anomalous chromomagnetic dipole moment of the top quark appears in
the top quark cross section, it is possible, due to uncertainties, to estimate the constraints that it would impose on the 
$\Delta \kappa$. For the  LHC, the
anomalous coupling is constrained to lie in the range $-0.09 \leq \Delta
\kappa \leq 0.1$   \cite{rizzo}.  Also the influence of an anomalous $\Delta \kappa$ on the cross section
and associated gluon jet energy for $t \bar t g$ has been analized, it leads to a
bound of $-2.1 \leq \Delta \kappa \leq 0.6$.

Anomalous couplings of the top quark to on-shell gluons would modify the rate for $B \to X_s \gamma$
\cite{us}. Using the recent data from CLEO collaboration for the
branching fraction of the process $B(b \to s\gamma)$ \cite{cleo}, we update the previous
analysis done in reference \cite{us} and get the new allowed region for the
anomalous chromomagnetic dipole moment of the top quark to be $-0.03 \leq \Delta
\kappa \leq 0.01$.

\begin{figure}[htbp]
  \centerline{\hbox{ \hspace{0.2cm}
    \includegraphics[width=6.5cm]{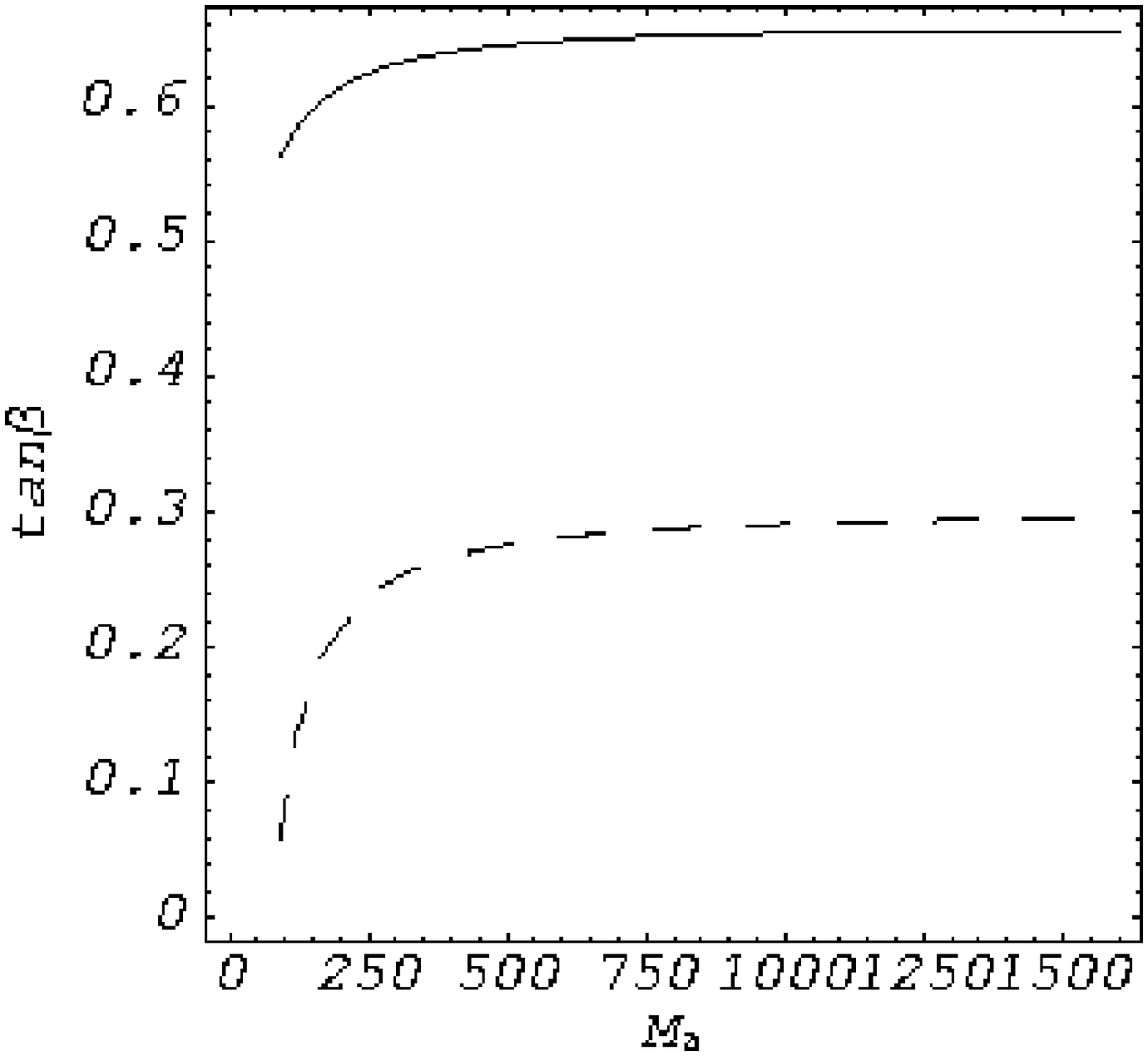}
    \hspace{0.3cm}
    \includegraphics[width=6.5cm]{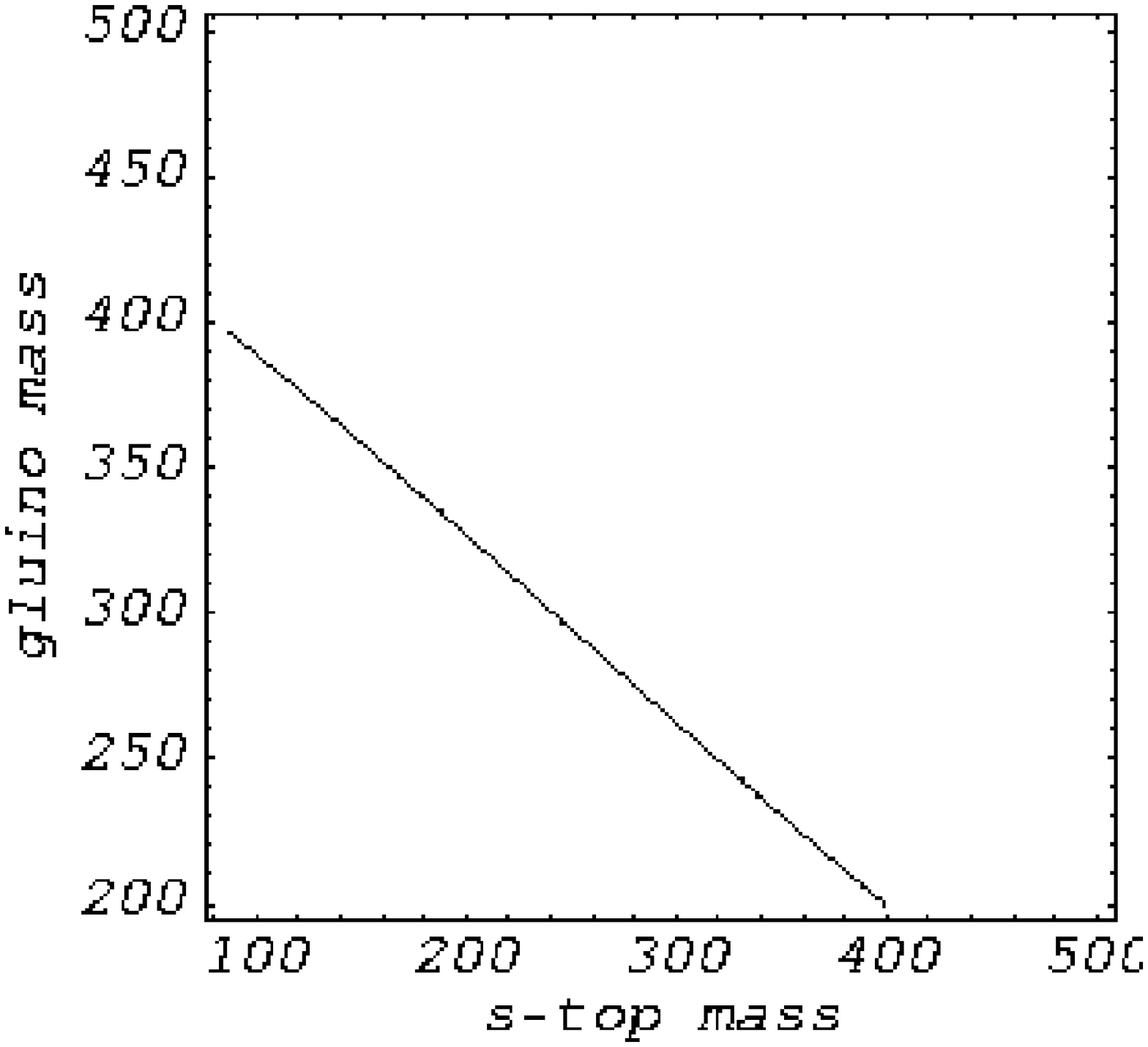}
         }}
 \caption{\it Contour plot for the plane  $\tan \beta-m_A$ with the common scalar mass $m_h=m_H=90(200)$ GeV 
for the solid line (dashed line). And Contour plot for the supersymmetric QCD contribution to the anomalous
chromomagnetic dipole moment of the top quark using the bound from the $b\to s \gamma$ process. }
\end{figure}
Beginning with the SM, the tipical QCD correction through gluon exchange
implies two different Feynman diagrams.
After the explicit calculation of the loops, the final result is,
\begin{equation}
\Delta \kappa=-\frac 16 \frac{\alpha_s(m_t)}{\pi} 
\end{equation} 
where we have a factor $-1/6$ coming from the color structure in the
diagram, $T^a T^b T^a=-T^b/6$ with $T^a$ the generators of $SU(3)_C$. The other possible contribution in the framework of the SM comes
from electroweak interactions. This contribution reads
\begin{equation}
  \Delta \kappa= - \frac{\sqrt{2} G_F m_t^2}{8
\pi^2} [H_1(m_h)+H_2(m_Z)] ,  
\end{equation}
where $H_i(m)$ are defined in reference \cite{us}.

The contributions within a general 2HDM will be different
from the SM contributions because of the presence of the virtual five
physical Higgs bosons which appear in any two Higgs doublet model after
spontaneus symmetry breaking: $H^0$, $A^0$, $h^0$, $H^\pm$ \cite{hunter}.
The expression  for the contribution of the neutral Higgs bosons is given by 
\begin{equation} \Delta \kappa=\frac{\sqrt{2} G_F}{8 \pi^2} 
[\lambda_{H^0 tt}^2 H_1(M_H^2)+\lambda_{h^0 tt}^2 H_1(M_h^2)+\lambda_{A^0
tt}^2 H_2(M_A^2)+\lambda_{G^0 tt}^2 H_2(M_Z^2)]
\end{equation}
where $\lambda_{itt}$ are the Yukawa couplings in the so-called models type I,
II and III \cite{hunter,us}. We show in figure 1 the allowed region (top-right)  for
the plane $\tan \beta$ vs $m_A$ with  $m_H=m_h=90(200)$ GeV for the solid (dashed) line. In this case we only find cuts with the upper limit for $\Delta \kappa$ from $b \to s \gamma$.  

Our last step is to calculate the anomalous chromomagnetic dipole moment
of the top quark in the framework of the MSSM. 
We only consider the SUSY QCD contribution which is generated from the
exchange of gluinos and sqaurks top. In this framework the SUSY QCD contribution arises from the exchange of  
$\tilde t_{1,2}$ and gluinos. The contribution for the virtual $\tilde t_1$ is given in ref \cite{us}.
Again, we plot the points which are of the same 
order of the $\Delta \kappa$ from $b \to s \gamma$ in the figure 1, right side.

In conclusion, the anomalous chromomagnetic dipole moment of the top quark gets the wide bound $\vert \Delta \kappa\vert \leq 0.45$ 
from the Tevatron experiments under the assumption that it is the only non-zero anomalous coupling. On the other hand, we found a more stringent 
bound from the transition $b \to s \gamma$, recently measured with improved precision by CLEO collaboration \cite{cleo}, and the bound is $-0.03 
\leq \Delta \kappa \leq  0.01$. Also, we have calculated the chromomagnetic dipole moment of the top
quark in the framework of the SM, 2HDM and SUSY-QCD. We found that in SM the
anomalous coupling is of the order of $\pm 10^{-2}$ and it is tending to $-0.004$ in the decoupling limit for the Higgs mass. Furthermore, 
the anomalous coupling is equal to $-7 \times 10^{-4}$ around a Higgs mass of $113$ GeV that is the lower experimental limit from LEP2.  
Now if we keep in mind the sensitivity of future and current experiments to $\Delta \kappa$ \cite{rizzo} a measurement of this anomalous coupling 
would suggest the presence of physics beyond the SM.  In the
2HDM the anomalous coupling of the top quark can reach $\sim 10^{-1}$ values, also gotten for the
supersymmetric QCD corrections. 

This work was supported by COLCIENCIAS, DIB and DINAIN.

\end{document}